\documentclass[twocolumn]{aastex63} 
\usepackage{xcolor} 
\usepackage{mathtools} 

\hypersetup{linkcolor=blue,citecolor=blue,filecolor=cyan,urlcolor=blue}

\newcommand{\ppm}{$\pm$} 
\newcommand{\ciii}{C\,\textsc{iii]}}
\newcommand{\civ}{C\,\textsc{iv}}
\newcommand{\niii}{N\,\textsc{iii]}}
\newcommand{\niv}{N\,\textsc{iv]}}
\newcommand{\nv}{N\,\textsc{v}}
\accepted{August 31,2021} 

\begin{document} 
\title{Evidence of a Tidal-disruption Event in GSN 069 from the Abnormal Carbon and Nitrogen Abundance Ratio} 
\email{shengzf@ustc.edu.cn, twang@ustc.edu.cn}

\author[0000-0001-6938-8670]{Zhenfeng Sheng}
\affiliation{CAS Key Laboratory for Research in Galaxies and Cosmology, University of Science and Technology of China, Hefei, Anhui 230026, China}
\affiliation{School of Astronomy and Space Science, University of Science and Technology of China, Hefei, Anhui 230026, China}

\author[0000-0002-1517-6792]{Tinggui Wang}
\affiliation{CAS Key Laboratory for Researches in Galaxies and Cosmology, University of Sciences and Technology of China, Hefei, Anhui 230026, China}
\affiliation{School of Astronomy and Space Science, University of Science and Technology of China, Hefei, Anhui 230026, China}

\author[0000-0003-4503-6333]{Gary Ferland}
\affiliation{ Department of Physics, University of Kentucky, Lexington, KY 40506, USA} 

\author[0000-0002-7020-4290]{Xinwen Shu}
\affiliation{Department of Physics, Anhui Normal University, Wuhu, Anhui 241002, China.}

\author[0000-0003-4975-2433]{Chenwei Yang}
\affiliation{SOA Key Laboratory for Polar Science, Polar Research Institute of China, 451 Jinqiao Road, Shanghai, 200136, China} 

\author[0000-0002-7152-3621]{Ning Jiang}
\affiliation{CAS Key Laboratory for Researches in Galaxies and Cosmology, University of Sciences and Technology of China, Hefei, Anhui 230026, China}
\affiliation{School of Astronomy and Space Science, University of Science and Technology of China, Hefei, Anhui 230026, China}

\author[0000-0002-3759-1487]{Yang Chen}
\affiliation{Anhui University, Hefei, Anhui 230601, China}
\affiliation{National Astronomical Observatories, Chinese Academy of Sciences, Beijing 100101, China}

\begin{abstract} 
GSN 069 is an ultra-soft X-ray active galactic nucleus that previously 
exhibited a huge X-ray outburst and a subsequent long-term decay. It has 
recently presented X-ray quasi-periodic eruptions (QPEs). We 
report the detection of strong nitrogen lines but weak or undetectable 
carbon lines in its far ultraviolet spectrum. With a detailed photoionization 
model, we use the \civ/\niv\ ratio and other ratios between  nitrogen 
lines to constrain the [C/N] abundance of GSN 069 to be from $-3.33$ to $-1.91$. 
We argue that a partially disrupted red giant star can naturally explain the abnormal C/N abundance in the UV spectrum, while the surviving core orbiting the black hole might produce the QPEs. 
\end{abstract} 
 
\section{Introduction} 
 
GSN 069 is an optically identified very low-mass bona-fide type 2 active 
galactic nuclei (AGN) with ultra-soft X-ray emission \citep{Miniutti2013}. 
It is the first case that presents X-ray quasi-periodic eruptions (QPEs) 
\citep{Miniutti2019Natur}. The X-ray emission of GSN 069 was first 
detected in 2010 and implied an outburst with a factor of more than 240 
increase in flux compared to the quiescence state \citep{Miniutti2013}. 
It has since shown a long-term decay \citep{Shu2018}. 

These observed 
properties were interpreted in several different scenarios. One possible 
interpretation links the outburst in 2010 to the AGN re-activation after 
a period of quiescence. In this case, the QPEs are driven by an accretion-flow 
instability, reminiscent of the heartbeat variability of black hole X-ray 
binaries \citep{Belloni1997, Altamirano2011,Miniutti2019Natur}. Another 
plausible scenario is that GSN 069 could experience a long-lived tidal-disruption event (TDE), manifesting a slow long decay of X-ray flux and 
spectral evolution \citep{Shu2018}. Alternatively, 
\cite{King2020MNRAS.493L.120K} suggested a stellar-mass white dwarf in a 
highly eccentric orbit about a massive black hole could reproduce the QPEs 
of GSN 069, while \cite{Ingram2021} proposed a self-lensing binary massive 
black hole interpretation. The last two works did not explain the long-term 
X-ray variations. Although similar X-ray QPEs have been detected in three more galaxies 
\citep{Sun2013, Giustini2020, Arcodia2021}, including two previously inactive ones, 
their origin remains a puzzle. 

Previous studies suggested that TDEs display a unique ultraviolet 
(UV) emission-line spectrum, characterized by strong nitrogen lines and weak carbon lines,  
distinct from those of AGNs \citep{Cenko2016,Yang2017,Brown2018}. 
This is intriguing because, in normal AGNs, the 
carbon emission lines like C \textsc{iii]} or C\textsc{ iv} are usually much stronger than 
nitrogen lines \citep{Vanden2001}, and even in rare nitrogen-rich (N-rich) 
AGNs there are significant carbon lines \citep[e.g., the most N-rich AGN 
Q0353-383,][]{Baldwin2003ApJ}. 
Enhancement in metallicity due to the stellar population  
has challenges in explaining the AGNs' N-rich phenomenon. 
TDEs can provide a more natural explanation because of  supplying a significant mass of N-rich/carbon-poor material \citep{Cenko2016, 
Kochanek2016MNRAS.458..127K}. Interestingly, examining the UV spectra of TDEs, \cite{Yang2017} suggested the \ciii$\lambda1908$/\niii$\lambda1750$ ratio is a good indicator of N/C 
abundance and derived abundance ratios for three TDEs which indicate 
nitrogen-enhanced core material of a disrupted star. Moreover, adopting this 
N/C abundance indicator, \cite{Liu2018} reported a candidate TDE in N-rich 
quasars \citep{Jiang2008,Batra2014} via abundance ratio variability.  

Inspired by these findings, we analyze the UV spectrum of 
GSN 069 to explore the possible clues to the question whether QPEs are 
directly associated with accretion-flow instabilities or due to extrinsic 
phenomena (e.g., such as a TDE or the presence of an orbiting body). Strikingly, 
we find that the nitrogen lines are anomalously strong, while the carbon 
lines (e.g., \ciii\ and \civ) are weak or nearly absent. We use the photoionization model to investigate the formation of nitrogen lines in GSN 069's UV spectrum, and explore the applicability of using the \civ/\niv\ 
line ratio as well as the ratios between different nitrogen lines to 
constrain the C/N abundance ratio.  

\section{UV Spectral decomposition} 
GSN 069 was observed twice spectroscopically in the far-UV (FUV) and 
near-UV (NUV) by the Space Telescope Imaging Spectrograph on board 
the Hubble Space Telescope, using the Multi-Anode Microchannel 
Array detector with the G140L and G230L gratings (Program ID: 13815 
and 15442, PI: Miniutti \citeyear{Miniutti2019Natur}). The first 
observation was performed in 2014 December and the second was in 2018 December. 
In each epoch, there were two 49 minute FUV and one 39 minute NUV exposures. We 
downloaded the pipeline-calibrated one-dimensional spectra from the 
Hubble Legacy Archive \footnote{https://hla.stsci.edu/hlaview.html}. 
After checking that there was no significant variability between the two 
epochs,  
the FUV and NUV spectra were signal-to-noise ratio (S/N) weighted and 
combined to a single spectrum covering the full UV range from 
1130$\sim$3100\,\AA. Then, we dereddened the Milky Way absorption 
according to the dust map provided by \cite{Schlegel1998}.  
 
\begin{figure} 
	\figurenum{1} 
	\label{fig:GSN069} 
	\centering 
	\epsscale{1.1} 
	\plotone{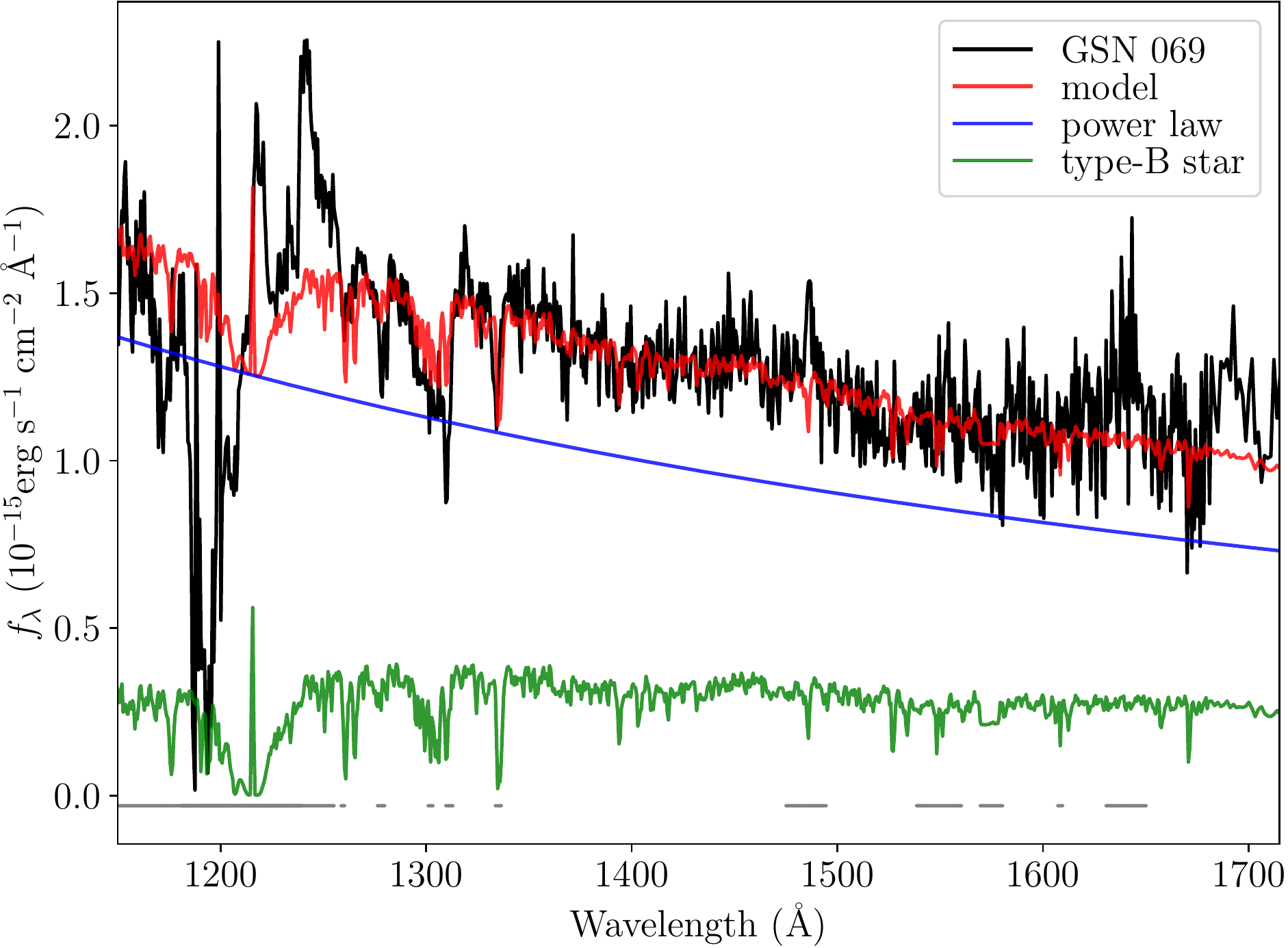} 
	\caption{Spectral decomposition of GSN 069. We show the UV spectrum 
of GSN 069 in black, and the best fitting model in red. The blue line 
represents the decomposed continuum, while the green spectrum represents 
the type-B-star template. The gray lines at the bottom represent the masked 
region when performing decomposition.} 
\end{figure} 
 
As mentioned by \cite{Miniutti2019Natur}, the UV spectrum of GSN 069 
resembles that of the main-sequence star (e.g., type-B star, see Extended 
Data Fig.2a of \citealt{Miniutti2019Natur}), indicating that it is likely 
contaminated by a relatively young stellar nuclear cluster. To remove the 
starlight contamination, we use intermediate early-type stars' co-added 
spectra (e.g., type-B0, -B3 and -B6 stars) from the Hubble Spectroscopic Legacy 
Archive 
(HSLA)\footnote{https://archive.stsci.edu/missions-and-data/hsla} to 
build a template stellar spectrum. We require that the spectra should have high 
S/N and cover the emission lines that we are interested in (e.g., \nv, \niv, 
\civ, \niii\ and \ciii). Also, no grating gaps should be present around these 
main emission lines. Only three stars satisfy our requirement, namely 
NGC1818-D1, EC05438-4741 and EC10500-1358, which are classified as B0-2 V-IV, 
B3-5 V-IV and B6-9.5 V-IV star, respectively (we mark them as T1, T2 and 
T3). However, only FUV data are available on the HSLA, so all 
the three stars' spectra did not cover the \ciii$\lambda$1909\,\AA. After 
resampling these spectra to match the resolution of GSN 069, their median 
S/N turned to 156.1, 78.7 and 101.1, respectively. We normalized the three 
spectra using the median flux between 1445 and 1455\,\AA\ to get the 
templates.  
 
Before the fitting process, we masked out regions of the prominent emission 
lines of GSN 069, strong Galactic absorption lines as well as geocoronal 
lines (see the bottom gray line in Figure \ref{fig:GSN069}). By adding a power-law continuum component to the mixed type B-star templates, 
the model can be written as, 
\begin{equation} 
f=A\times(\lambda/1000)^{\alpha}+a\times T_1+b\times T_2+c\times T_3 
\end{equation}  
where (A,$\alpha$,a,b,c)=(1.71\ppm0.04, $-$1.57\ppm0.05, 0.04\ppm0.02, 0, 
0.29\ppm0.02). 
 
\begin{figure*} 
	\figurenum{2} 
	\label{fig:Lines} 
	\centering 
	\epsscale{1.2} 
	\plotone{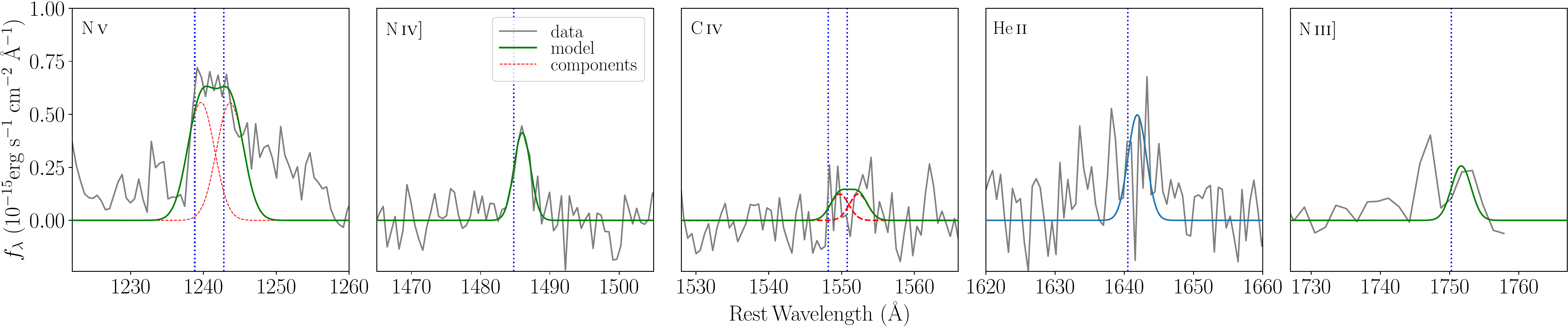} 
	\caption{Emission lines of GSN 069. In each panel, we show the 
starlight and continuum subtracted spectrum in gray, and we use the 
vertical dotted lines to mark the wavelength of the emission lines. The 
blue and red lines represent the emission line models and components, 
respectively.} 
\end{figure*}

Thus, the continuum and the B-star components can be removed from the 
original spectrum to get the emission lines of GSN 069 (see Figure 
\ref{fig:Lines}). We use one Gaussian to model the single line (\niv\ and 
He\,\textsc{ii}) and two Gaussian's to model the doublets (\nv, \niv\ and \civ). Considering 
the \niv\, has the best profiles and S/N, we firstly fit the \niv. With the 
best fitted \niv, the profiles of \civ, He\,\textsc{ii} and \niii\ are assumed to be 
the same as \niv. In order to estimate the uncertainties, we generate 100 
mock spectra by adding Gaussian noise to the spectrum using the uncertainty 
of input data. By repeating the fit procedure, the standard deviation of 
the distributions of best fit is taken as the uncertainty for each spectral 
quantity. 
 
The fitted line flux of \nv, \niv, \civ, He\,\textsc{ii}, \niii\ are 5.16\ppm0.19, 
1.24\ppm0.19, 0.50\ppm0.19, 1.65\ppm0.27 and 0.91\ppm0.31 (in unit of 
$10^{-15} \rm erg/s/cm^2$), respectively. 
Although our B-star templates do not cover \ciii, we examined the 
theoretical stellar spectra of B stars \citep{Castelli2003IAUS,Castelli2005MSAIS} and found 
no strong absorption around the \ciii\ emission lines. We fit a Gaussian and 
a local continuum to spectrum in \ciii\ window. The line is completely not 
detected. 
 
\section{Constraint on the C/N ratio} 
\begin{figure*} 
	\figurenum{3} 
	\label{fig:LP} 
	\centering 
	\epsscale{0.95} 
	\plotone{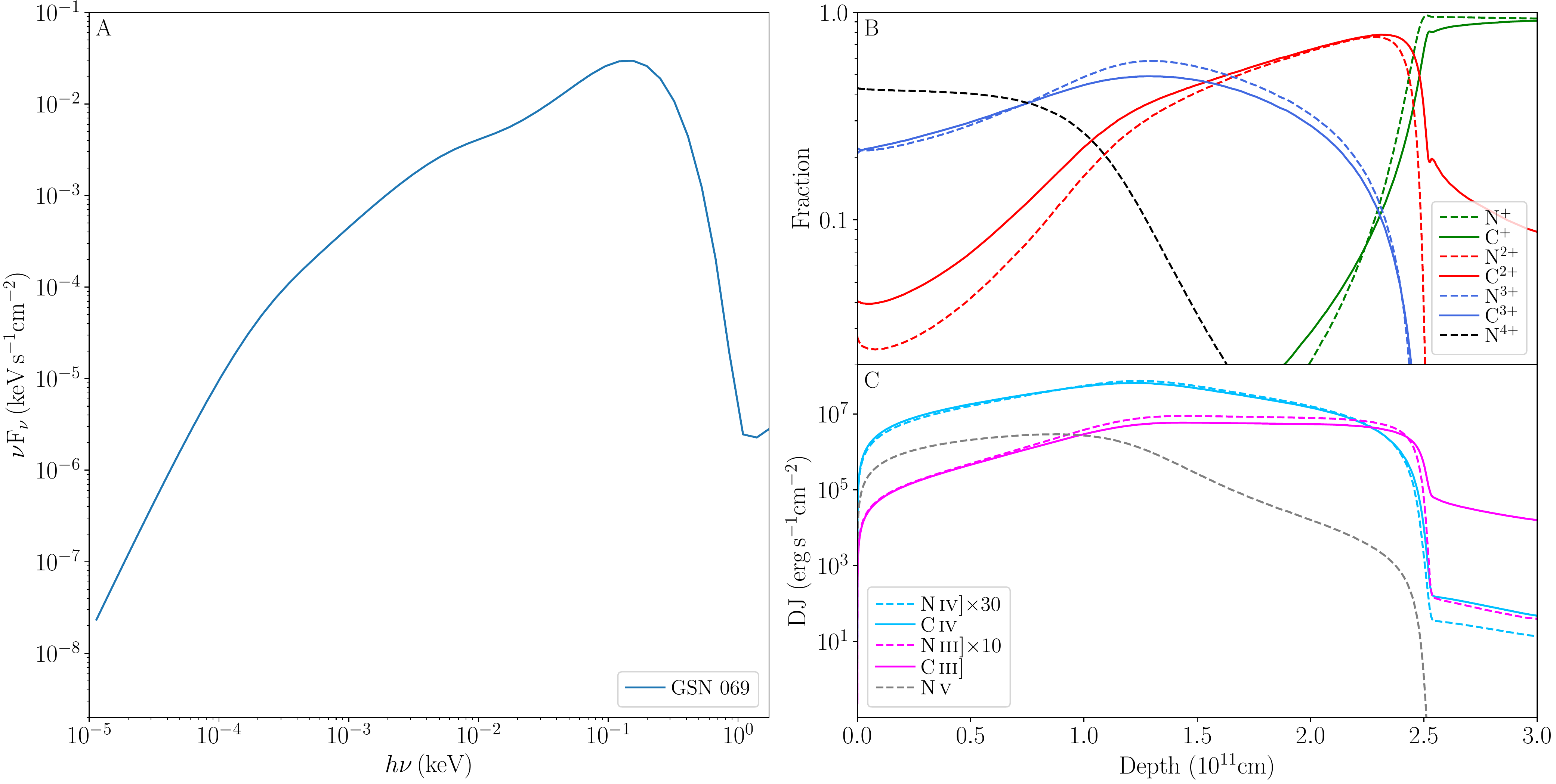} 
	\caption{Panel (A) presents the SED of GSN 069. Panel (B) and (C) 
present ionization fractions and depth-weighted line emissivities against 
spatial depth into a cloud having $U=-0.2$, $n_{\rm H}=10^{10}\rm cm^{-3}$, 
solar abundances, and an incident SED of GSN 069. The dashed curves 
represent nitrogen and the solid curves represent carbon.} 
\end{figure*} 
 
\begin{figure*} 
	\figurenum{4} 
	\label{fig:C4N4} 
	\centering 
	\plotone{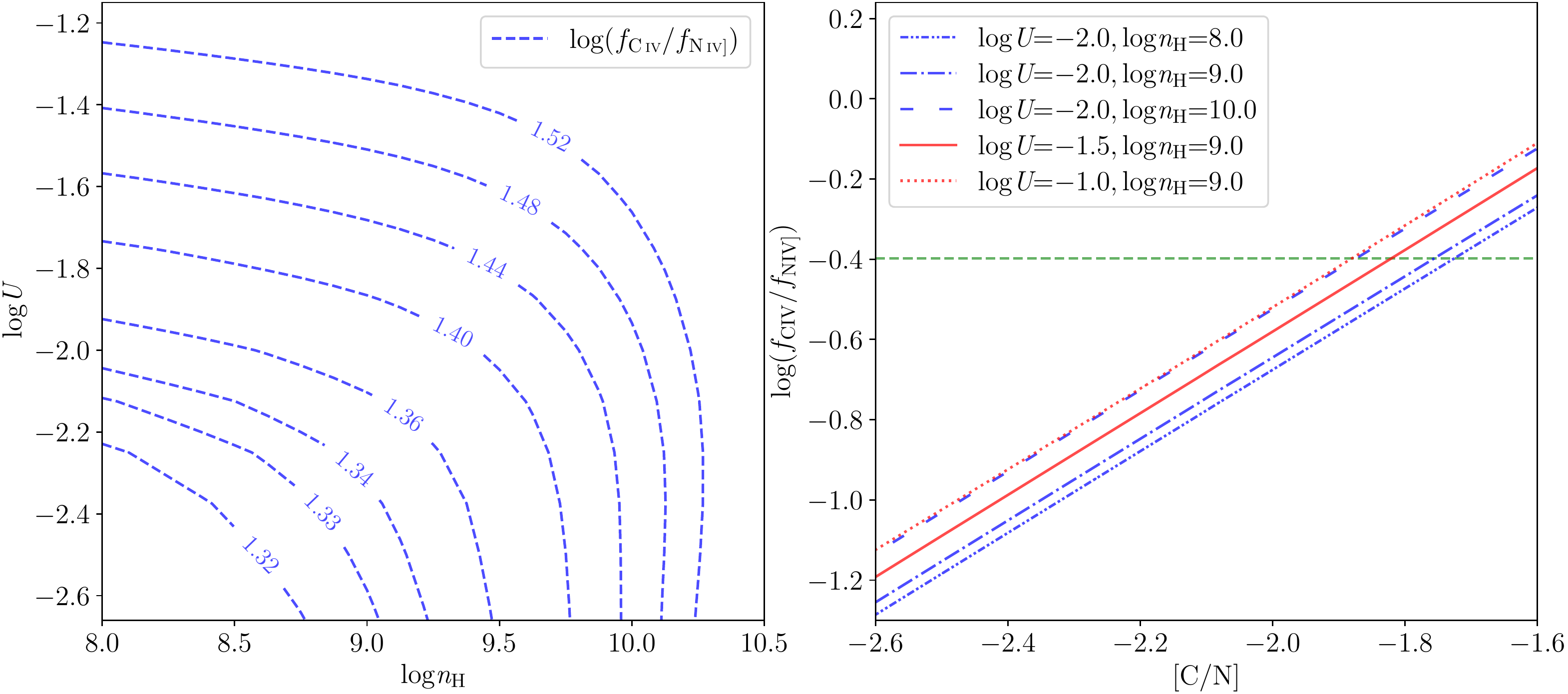} 
	\caption{Left panel: contour plots for the logarithmic line ratio of 
\civ/\niv\ in the $U\sbond n_{\rm H}$ plane at solar metallicity. Right panel: 
theoretical logarithmic line ratios of  \civ/\niv\ as a function of [C/N] in 
different $U$ and $n_{\rm H}$. The horizontal green dashed line marks the 
observed  \civ/\niv\  ratio of GSN 069.} 
\end{figure*} 
 
\begin{figure*} 
	\figurenum{5} 
	\label{fig:AR} 
	\centering 
	\plotone{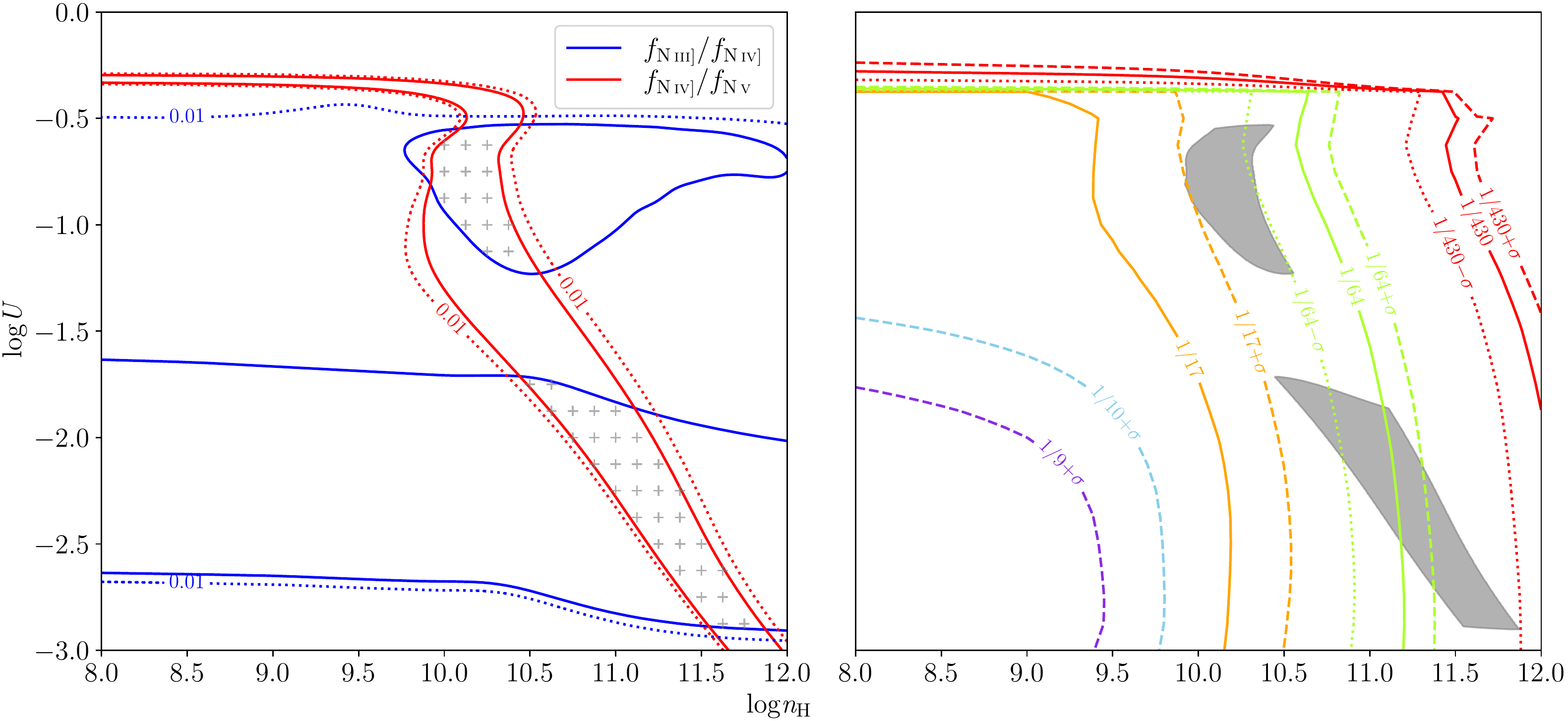} 
	\caption{Left panel: contour plots for the allowed region in $U\sbond n_{\rm H}$ plane defined by the $f_{\rm \niii}/f_{\rm \niv}$ (blue) and $f_{\rm \niv}/f_{\rm \nv}$ (red), respectively. The solid and dotted lines represent the 95\% and 99\% confidence level of line ratios, respectively. The gray cross symbols roughly mark the enclosed area jointly defined by the $f_{\rm \niii}/f_{\rm \niv}$ and $f_{\rm \niv}/f_{\rm \nv}$ (taking the 95\% confidence level, respectively). Right panel: the gray shadow area is the same allowed region jointly defined by the $f_{\rm \niii}/f_{\rm \niv}$ and $f_{\rm \niv}/f_{\rm \nv}$ as marked in the left panel. The solid line contours are $f_{\rm \civ}/f_{\rm \niv}$ at the observed value for different carbon abundance, while the dashed and dotted lines represent the corresponding 1$\sigma$ upper limit and 1$\sigma$ lower limit of the observed value (marked with the additional label $+1\sigma$ and $-1\sigma$), respectively. From lower left to upper right, the carbon abundance decreasing from $Z_{C,\sun}$/9 to $Z_{C,\sun}$/430. The models with 1/17 to 1/430 solar carbon abundances intersect the allowable shadowed gray region.
	} 
\end{figure*}

Our goal is to constrain C/N ratio of GSN 069 from its observed UV line 
ratios. Previous works have used the ratios of nitrogen emission lines to 
various collisional/recombination lines of other elements to derive the 
metal abundance in the broad line region (BLR) of quasars based on the chemical 
enrichment that nitrogen is produced by a secondary process 
(\citealt*{Hamann1993,Hamann1999}; \citealt{Hamann2002}; 
\citealt{Batra2014}). \cite{Yang2017} showed that \ciii/\niii\ ratio is a 
good indicator of [N/C] abundance ratio that is insensitive to the shape 
of the ionizing continuum or the physical parameters of the emission line 
region over wide parameter ranges. Accordingly, they set a lower limit on 
the N/C ratio for three TDEs. Because \ciii\ is not detected and \niii\ is 
only marginally detected in GSN 069, it is not possible to employ the same 
method. However, \niv\ and \nv\ are prominent in GSN 069, so we will use the 
\civ/\niv\ and ratios of between different nitrogen lines to constrain the C/N abundance ratio and the physical conditions of the line emission region. 
Although a large number of grid models already exist in the literature for 
the conditions of BLR in quasars 
\citep{Korista1997ApJS..108..401K}, more precise modeling requires a 
good match in the input ionizing continuum to that of GSN 069 and the 
specific chemical abundance. Thus we run a set of photoionization models 
using CLOUDY 17.02 \citep{Ferland2017} by taking these into consideration.   
 
We adopt an SED derived by fitting a disk-corona model to the observed data 
from optical UV to X-ray \citep{Miniutti2013}, which is shown in the left panel of Figure 
\ref{fig:LP}. Although it is still unclear whether there are cold gas clouds on a BLR scale in the nuclei of low-luminosity AGNs or quiescent galaxies, some theoretical and observational works suggest that BLR disappears when the Eddington ratio of an AGN is below 0.001 (e.g., \citealt{Elitzur2009}). The abnormal [N/C] in two of three TDEs analyzed by \cite{Yang2017}, suggests that they are unlikely normal ISM, even considering the high metallicity in galactic nuclei \citep{Nagao2006}. As for the TDE, the disrupted debris, either in outflow or accretion disk, is predicted to be photoionized and produce optical emission lines (\citealt{Roos1992}, \citealt{Bogdanovic2004}, \citealt{Strubbe2009}). The gaseous debris of a star can be high density and may well have an unusual chemical composition \citep{Cenko2016, Kochanek2016MNRAS.458..127K}. In analog with an analysis of BLR of AGNs (\citealt{Peterson1997}, p. 73), only $\sim10^{-3}\rm M_{\sun}$ line emitting gas is required to produce the emission lines. So in the context of the tidal-disruption model, the line emitting 
gas is assumed to come from the interior of a star that is polluted with 
nuclear fusion production.

For main-sequence stars, which account for the 
majority of observed TDEs, most carbon in the center of the star is 
converted into nitrogen in a relatively short time in the CNO cycle. In 
equilibrium, the C/N ratio is a function of temperature. We set the overall 
abundance to solar values given by \cite{Grevesse2010}, but the allow carbon 
and nitrogen ratio to vary while the sum of carbon and nitrogen is maintained to the solar value according to the CN cycle. Then a set of plane-parallel slab models with 
constant density and ionization parameter are computed. The logarithm of gas density ($n_{\rm H}$) in cm$^{-3}$ runs from 8 to 12, and the logarithm 
of ionization parameters (${\rm U}=\Phi/4\pi r^2 n_{\rm H} c$, where 
$\Phi$ is the flux of ionizing photons and $c$ is the speed of light) from -3 
to 0. Both the $n_{\rm H}$ and $U$ are run with an initial step of 0.25 
dex to estimate the range of parameters and a fine grid of models with a 
step of 0.125 dex around plausible parameter regimes are computed later 
to determine more precisely the physical conditions and the upper limit 
of carbon abundance.  
 
In the panel (B) of Figure \ref{fig:LP}, we present the ionization structure 
of a cloud having $\rm log \emph U=-0.2$, $n_{\rm H}=10^{10}\rm cm^{-3}$, 
solar abundances, and an incident SED of GSN 069, while the depth-weighted 
line emissivities against spatial depth are shown in the panel (C). The 
emission regions of \civ\ and \niv\ significantly overlap as well as that of 
\ciii\ and \niii, indicating these pairs are also good abundance indicators 
(similar plots were also presented by \citealt{Hamann2002}). We firstly examine 
the dependences of the line ratio of \civ/\niv\ (herein after, $f_{\rm \civ}/f_{\rm \niv}$) upon the $U$, $n_{\rm H}$ and [C/N] abundance. In the left panel of Figure \ref{fig:C4N4}, we plot the contours 
of logarithmic $f_{\rm \civ}/f_{\rm \niv}$ in the $U\sbond n_{\rm H}$ plane 
at solar abundance as an example. At the given [C/N] abundance and $U$, 
when the density is lower than the critical density of \niv\  
($n_c\sim10^{10}\rm cm^{-3}$), the $f_{\rm \civ}/f_{\rm \niv}$ is not sensitive to the 
density (e.g., when $n_{\rm H}$ varied $\sim$1 dex, the ratio just changes 
$\sim$0.04 dex). However, while the $n_{\rm H}$ is approaching the critical 
density, the contours drop significantly due to collisional de-excitation 
of \niv\ is overwhelming. The $f_{\rm \civ}/f_{\rm \niv}$ also has a weak dependence with 
$U$. For example, by fixing the $n_{\rm H}=8.0$, when the $\rm log \emph 
U$ changes from $-2.2$ to $-1.2$ (1 dex), the ratio just changes $\sim$0.2 dex. 
This weak dependence on density or ionization is much clearer in the right 
panel of Figure \ref{fig:C4N4}, in which we plot theoretical logarithmic 
$f_{\rm \civ}/f_{\rm \niv}$ as a function of [C/N] in different $\rm 
log\emph{U}$ and $\rm log \emph n_{\rm H}$. We can see that the $f_{\rm \civ}/f_{\rm \niv}$ significantly depends on the [C/N]. A similar plot is also 
presented by \cite{Yang2017} who used the \ciii/\niii\ ratio against [C/N] 
to constrain the carbon abundance of three TDEs. Thus, following 
\cite{Yang2017}, we can also estimate the carbon abundance of GSN 069 by 
drawing the observed $f_{\rm \civ}/f_{\rm \niv}$ on the right panel of Figure 
\ref{fig:C4N4} (horizontal green dashed line), from which we infer 
$[\rm C/N]\sim-1.88$ by roughly taking  $\rm log \emph U=-2$ and $\rm log \emph 
n_{\rm H}=10$.

To set the plausible ranges of density and ionization parameters more 
precisely, we further examine the line ratios of $f_{\rm \niii}/f_{\rm 
\niv}$ and $f_{\rm \niv}/f_{\rm \nv}$ and find they are not sensitive to 
the C/O ratios. In the left panel of Figure \ref{fig:AR}, we present the 
contours of $f_{\rm \niii}/f_{\rm \niv}$ (blue) and $f_{\rm \niv}/f_{\rm 
\nv}$ (red) in the $U\sbond n_{\rm H}$ plane. The solid and dotted lines 
represent the 95\% and 99\% confidence level of line ratios, respectively, 
by assuming that emission line ratios follow normal distributions with mean 
and sigma from measurements. The gray cross symbols roughly mark the 
enclosed area jointly defined by the 95\% confidence level of $f_{\rm 
\niii}/f_{\rm \niv}$ and $f_{\rm \niv}/f_{\rm \nv}$, which is extracted and 
re-plotted in the right panel of Figure \ref{fig:AR}.  We have verified 
that carbon abundance has a negligible effect on these contours.  
 
Next, we add the contours of the ratio of \civ\ to \niv\ at the observed value 
for different C abundance in the right panel of Figure \ref{fig:AR}. The 
inline labels represent the fraction of solar carbon abundance. The solid line contours are the ratio of \civ\ to \niv\ at the observed value, while the dashed and dotted lines represent the corresponding 1$\sigma$ upper limit and 1$\sigma$ lower limit (marked with the additional label $+1\sigma$ and $-1\sigma$), respectively. Models with carbon abundance at 1/10 of the solar value or higher cannot produce a 
\civ/\niv\ ratio ($f_{\rm \civ}/f_{\rm \niv}$) within the observed values for the range of physical parameters, so they do not appear in the figure; neither does the 1$\sigma$ lower limit of $f_{\rm \civ}/f_{\rm \niv}$ at 1/17 of solar carbon abundance. However, the models can produce the 1$\sigma$ upper limit of the observed $f_{\rm \civ}/f_{\rm \niv}$ at $Z_{C,\sun}/9$ or lower abundance.

With the decrease in carbon abundance, the regime of $f_{\rm \civ}/f_{\rm \niv}$ shifting from lower left to upper right and the models with 1/17 to 1/430 solar carbon abundances intersect the allowable region (shadow gray region). Thus we obtain an upper limit of the carbon abundance at $Z_{C,\sun}/17$ and a lower limit at $Z_{C,\sun}/430$.

\section{Discussion} 
According to the above analysis, we constrain the carbon abundance of GSN 
069 in the range from $Z_{C,\sun}/430$ to $Z_{C,\sun}/17$, and the 
corresponding [C/N] ranges from $-3.33$ to $-1.91$. 
We have assumed that the emission lines come from the same region. In panels 
(B) and (C) of Figure \ref{fig:LP}, these lines do overlap substantially. 
However, it is plausible that the emission line region is stratified as 
in Seyfert galaxies, so \niii\ and \nv\ may come from different regions. With 
a similar ionization potential, \niv\ and \civ\ is likely formed in the same 
zone, so they can be used as a good indicator of abundance. In that case, 
\niii/\niv\ defines a lower limit on the ionization parameter of \niv\ 
emission region, so the contours of \niii/\niv\ (blue) in the left panel 
of Figure \ref{fig:AR} should shift upward. Likewise, we overestimate the 
\niv/\nv\ from \niv\ emission region, thus contours of \niv/\nv\ (red) in the 
left panel of Figure \ref{fig:AR} should shift to the lower left. By considering this effect, $Z_{C,\sun}/9$ is still a safe upper limit of carbon 
abundance, yielding $\rm[C/N]<-1.61$, which corresponds to a N/C ratio larger 
than 10.23. In this instance, the N/C abundance of GSN 069 is still extremely 
abnormal compared with that of the reported three TDEs \citep{Yang2017}.  
 
\cite{Kochanek2016MNRAS.458..127K} suggested that the tidal-disruption of a 
main-sequence star can naturally explain the N-rich phenomenon for the 
rapid enhancement of N and the depletion of C in the CNO cycle. Indeed, 
\cite{Shu2018} argued that the X-ray outburst of GSN 069, the consequent 
flux decay and ultra-soft X-ray spectrum could be attributed to a 
long-lived TDE, while they cannot completely rule out the highly variable AGN activity. 

The extreme anomalous N-rich nature we found in the UV spectra of GSN 069 provides strong evidence that a TDE has occurred in this AGN. The disrupted star is probably a red giant star with an inert compact helium core surrounded by a shell of hydrogen 
fusing via the CNO cycle. When the star enters the red giant branch, the stellar envelope is less gravitationally bound than it was in the main-sequence phase, so the star could be partially disrupted, leaving the dense stellar core. In this scenario, the accreted material should be N-rich and the surviving core may still orbit the black hole producing QPEs \citep{King2020MNRAS.493L.120K}.

We note that GSN 069 is a type 2 AGN, so the possibility that the QPEs are caused by an AGN activity, such as the disk instability model \citep{Miniutti2013,Miniutti2019Natur}, cannot be fully ruled out. However, \cite{Arcodia2021} report two further QPEs occurred in quiescent galaxies, indicating that AGN activity is not required to trigger these events, as the two host galaxies have not been active for at least the last $10^{3}\sim10^{4}$ years. This in turn further supports our model that QPEs of GSN 069 are triggered or driven by a total or partial TDE \cite{King2020MNRAS.493L.120K} or some other extreme mass ratio inspiral scenario \citep{Arcodia2021,Metzger2021}.

\section{Summary} 
In this work, using the photoionization model, we find that \civ/\niv\ is 
not as sensitive to the ionization parameter or gas density as it is to 
the C/N abundances and can be taken as an abundance indicator. By 
investigating the UV spectrum of GSN 069, we use the \civ/\niv\ ratio as well 
as the ratios between different nitrogen lines to constrain its C/N 
abundance. We report that the carbon abundance of GSN 069 ranges from 
$Z_{C,\sun}/430$ to $Z_{C,\sun}/17$, and the corresponding [C/N] is from 
$-3.33$ to $-1.91$. Such extreme anomalous N-rich and C-poor phenomena can 
be naturally attributed to a TDE. The results suggest that GSN 069 experienced a 
partial disrupted red giant star which caused the previous X-ray outburst, 
consequent flux decay, and the abnormal C/N abundance in the UV spectrum, 
while the surviving core orbiting the black hole might drive the QPEs. 
\\ 
\\
\\

We thank the anonymous referee for valuable comments and
suggestions that lead to significant improvement in the quality of
the paper. This research is supported by National Natural Science Foundation of China (NSFC-11833007, 12103048, 11822301, 12073025, 12003001) and Fundamental Research Funds for the Central Universities. The data used in this research are based on observations made 
with the NASA/ESA Hubble Space Telescope, and obtained from the Hubble 
Legacy Archive, which is a collaboration between the Space Telescope Science Institute (STScI/NASA), the Space Telescope European Coordinating 
Facility (ST-ECF/ESA) and the Canadian Astronomy Data Centre 
(CADC/NRC/CSA). This research has made use of the HSLA database 
\citep{HSLA2017}, developed and maintained at STScI, Baltimore, USA. This 
research made use of Astropy, a community-developed core Python 
package for Astronomy \citep{Astropy2013} and PyAstronomy, a collection 
of astronomy related packages \citep{pya}.  
 
\bibliographystyle{aasjournal}

\end{document}